\documentclass[%
 reprint,
 amsmath,amssymb,
 aps,
 superscriptaddress,
onecolumn,
longbibliography]{revtex4-2}

\usepackage{graphicx}
\usepackage{dcolumn}
\usepackage{bm}
\usepackage{epsfig}

\usepackage{color}
\usepackage{graphicx}
\newcommand{\jad}{\textcolor{blue}}

\begin{document}

\preprint{APS/123-QED}

\title{Darcy-Reynolds forces during intrusion into granular-fluid beds}

\author{Joshua Strader}
\author{Neil Causley}%
\affiliation{Department of Physics, Naval Postgraduate School, 833 Dyer Road, Monterey, CA 93943}


\author{Joshua A. Dijksman}
\affiliation{Physical Chemistry and Soft Matter, Wageningen University and Research, Stippeneng 4, 6708 WE Wageningen, Netherlands}
\author{Abram H. Clark}
\affiliation{Department of Physics, Naval Postgraduate School, 833 Dyer Road, Monterey, CA 93943}


\date{\today}

\begin{abstract}
We experimentally study intrusion into fluid-saturated granular beds by a free-falling sphere, varying particle size and fluid viscosity. We test our results against Darcy-Reynolds theory, where the deceleration of the sphere is controlled by Reynolds dilatancy and the Darcy flow resistance. We find the observed intruder dynamics are consistent with Darcy-Reynolds theory for varied particle size. We also find that our experimental results for varied viscosity are consistent with Darcy-Reynolds theory, but only for a limited range of the viscosity. For large viscosities, observed forces begin to decrease with increasing viscosity, in contrast with the theoretical prediction.
\end{abstract}

\maketitle


\section{\label{sec:intro}Introduction}

Intrusion or impact into a granular-fluid mixtures is a common process with relevance in, e.g., bio-inspired locomotion problems~\cite{Nordstrom2015,kudrolli2019burrowing} or shock absorption applications~\cite{gurgen2017shear}. Each of the constituent phases of this system (granular flows or fluid flows) is already challenging enough to describe, and the combination is even more difficult. There has been extensive recent work on steady-state rheology of granular fluid mixtures, including generalizations of inertial rheology descriptions for granular flows~\cite{Boyer2011,Trulsson2012,Guazzelli2018,Pahtz2019} as well as rheological studies of shear-thickening behavior~\cite{Brown_2014,cates,seto2013discontinuous}. Steady state analyses of granular and suspension flows often assume a weak coupling between the dynamics of both phases or only mild gradients (space or time) in the flow rate or the local stress. These analysis therefore offer few handles to understand more complex flow situations such as intrusion, which inherently involves compaction or dilation and propagation phenomena~\cite{Umbanhowar2010, waitukaitis2012impact,clark2015,han2016high}. Thus, in addition to the utility of describing the impact or intrusion process for practical applications, intrusion is a useful benchmark to probe the limits of existing theories and uncover new physics. This has been recognized by a diverse and expanding set of works on intrusion~\cite{goldman2008,waitukaitis2012impact,peters2014quasi,vandermeer2017impact}. In particular, understanding how relevant data during an impact (e.g., crater size~\cite{Walsh2003,Uehara2003}, or peak forces~\cite{goldman2008,Krizou2020}) depends on system parameters (e.g., intruder speed, intruder size, or grain stiffness) often yields significant insight about underlying physics, especially inherently transient processes that by definition cannot be captured by steady-state descriptions.

A notable example of such a process was recently highlighted by~\citet{jerome2016}. When intrusion occurs into a granular bed in which the packing fraction $\phi$ is compacted above a critical volume fraction $\phi_c$ (due to, e.g., external vibrations~\cite{nowak1997reversibility,Pugnaloni2008} or aging from other mechanisms~\cite{gago2020universal,Allen2018}), there is an initial transient where the bed dilates, due to Reynolds dilatancy~\cite{reynolds}. Generally, the bed will be saturated in some fluid (e.g., air or water), and the fluid will be sucked into the expanding pore structure. For sufficiently small particles and a sufficiently viscous fluid, Darcy pressure~\cite{darcy} begins to play an increasingly dominant role during the granular bed expansion. \citet{jerome2016} formulated a basic theory that combined these two effects, called Darcy-Reynolds theory (DRT), to describe the dynamics of spheres impacting granular beds that were fully immersed in a fluid. The authors showed explicitly that DRT could explain the dependence on $\phi$ of the force response during intrusion into fluid-saturated granular beds.


Although the $\phi$-dependence of the impact response was confirmed to agree with DRT~\cite{jerome2016}, several other parameters like viscosity $\eta_f$ of the interstitial fluid or the particle diameter $d$ play a crucial role in this theory, but the scaling behavior for these parameters was tested only for a few cases. If tested and confirmed, this would provide a framework that could be used for, e.g., prediction of robotic locomotion behavior~\cite{agarwal2021surprising} or tunable granular-fluid mixtures. One particular example motivating this study is the use of a ferrofluid as the viscous fluid. Ferrofluids~\cite{raj1995advances} consist of nanometer-sized iron particles coated in a surfactant suspendend in a simple fluid (e.g., alcohol or a petroleum-based fluid). Ferrofluids behave approximately as viscous fluids, but with a viscosity that depends on the applied magnetic field. Thus, the viscosity can be changed \textit{in situ} during some deformation of the material to achieve desired results. Since Darcy-Reynolds pressure increase with increasing $\eta_f$, this could provide a way to externally tune the flow behavior of granular-fluid mixtures.

\begin{figure}[!t]
    \centering
    \includegraphics[width=\columnwidth]{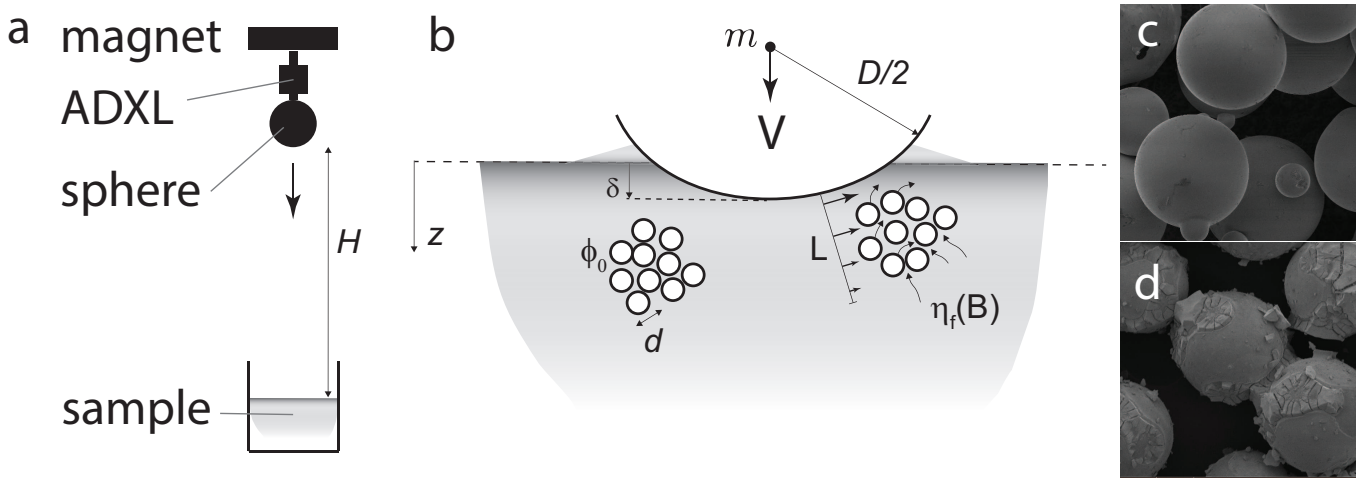}
    \caption{(a) Sketch of the experiment. The magnet releases the sphere, which is connected to the accelerometer (ADXL) via a threaded rod. The magnet is dropped from height $H$ onto the submersed settled particle bed (sample). (b) Sketch of the mechanism slowing down the spherical intruder impacting the bed of particles with diameter $d$ and packing fraction $\phi_0$. Under influence of the penetration of the intruder with speed $V$, diameter $D$ and mass $m$ the particles will bulge out from the bed surface (long dashed line). At penetration depth $z=\delta$ the particle have been sheared in a region with size $L$. This shear required dilating the packing and thus the local absorption of fluid with viscosity $\eta_f(B)$ whose properties may depend on the applied magnetic field strength $B$. (c) SEM image of the glass beads used, showing slight polydispersity and a mild roughness (d) SEM image of dried ferrofluid on the glass beads. The individual ferrofluid nanoparticles are too small to observe yet the leftovers from capillary bridges are clearly visible. Image width in (c,d) is 200 micrometer.}
    \label{fig:sketch}
\end{figure}

Here, we demonstrate how DRT predicts scaling laws for intrusion into fluid-grain mixtures as a function of particle size and fluid viscosity. We then experimentally test these scaling laws with impact experiments by dropping spheres from a height $H$ into fluid-saturated granular beds with varying particle size $d$ and fluid viscosity $\eta_f$; see Fig.~\ref{fig:sketch}(a). In both cases, we find results that are consistent with DRT over a range of parameter values. We observe some change to the phenomenology for large $d$; we show that this can be explained using DRT. However, we also observe a qualitative discrepancy for the predicted behavior for large $\eta_f$ which cannot be reconciled with DRT. Increasing $\eta_f$ should lead to increasing Darcy-Reynolds forces. Instead, we observe that for large $\eta_f$, increasing $\eta_f$ leads to decreasing forces during impact. Viscosity is controlled by adding glycerol to water as well as by using a ferrofluid and tuning the viscosity with an external applied magnetic field; both methods yield similar results. Our results demonstrate that Darcy-Reynolds theory as formulated describes variation in $d$ over a wide range but breaks down for large viscosities, at least for the particles and fluids studied here. However, the overall agreement between the glycerol-water mixtures and the ferrofluid suggest that ferrofluids could be used to construct tunable complex fluids, where the applied magnetic field controls the viscosity and hence impact hardness.


\section{\label{sec:theory} Darcy-Reynolds Theory}

We first reproduce the derivation of Darcy-Reynolds theory that appears in the main text and the Supplemental Material of~\citet{jerome2016}. The key idea is that intrusion of an object requires shear in the particulate phase. Such shear gives rise to frictional stress as set by some effective friction coefficient. Additionally, when a packing is denser than its ``critical state'' solid fraction $\phi_c$---i.e. the density it would have during steady-state shear---then the material will dilate when sheared until asymptotically approaching $\phi_c$, a process known as Reynolds dilatancy. This induces fluid flow into the expanding pores. Under certain conditions the pore fluid pressure $P_f$ can locally reach values much larger than any other local pressure scale, such as gravity, in which case $P_f$ dominates the dynamics. The key hurdle is then to find an expression for $P_f$ for the situation sketched in Fig.~\ref{fig:sketch}(b). 

It is reasonable to assume that the rate of dilation is linearly proportional to the shear rate $\dot{\gamma}$~\cite{sakaie2008mr}. Thus, the first assumption of this theory is that the dilation obeys a simple differential equation,
\begin{equation}
    \frac{1}{\phi}\frac{\partial \phi}{\partial t} = -\alpha \dot{\gamma} (\phi - \phi_c).
    \label{eqn:Reynolds}
\end{equation}
To approximate the magnitude of $\alpha$, we take finite differences, i.e., $\dot{\gamma} \rightarrow \delta \gamma / \delta t$. We assume that the strain needed over which dilation happens is $\delta \gamma \approx 0.1$~\cite{sakaie2008mr,Kabla2009,vasisht2020}, and $\partial \phi / \partial t \rightarrow \delta \phi / \delta t$, with $\delta \phi \approx \phi_c - \phi$, yielding $\alpha = 1 / (\phi \delta\gamma)$. If $\phi \approx 0.6$, then $\alpha \approx 20$. We use this approximation later to confirm that our comparison between theory and experiments is reasonable. 

As previously stated, when the granular phase dilates, fluid from elsewhere in the material must fill this volume opened up via dilation. If the particle diameter $d$ is small, then the Darcy flow resistance of the fluid through the pore structure is dominant. The Darcy law states that
\begin{equation}
    (1-\phi)(\mathbf{V}_f-\mathbf{V}_p) = -\frac{\kappa}{\eta_f}\nabla P_f,
    \label{eqn:Darcy}
\end{equation}
where $\kappa \propto d^2$ is the permeability, $\eta_f$ is the fluid viscosity, and $P_f$ is the pore pressure. Assuming that the particle and fluid phases are both incompressible, i.e., $\partial \phi / \partial t + \nabla \cdot (\phi \mathbf{V_p}) = 0$ and $\partial \phi / \partial t + \nabla \cdot (\phi \mathbf{V_p}) = 0$, then taking the divergence of Eq.~\eqref{eqn:Darcy} yields $(1/\phi)\partial\phi/\partial t = -(\kappa^2/\eta_f)\nabla^2 P_f$, assuming that spatial dependence in $\phi$ can be neglected (i.e., $\nabla \phi =0$). Combining with Eq.~\eqref{eqn:Reynolds} yields
\begin{equation}
    \nabla ^2 P_f  = \frac{\eta_f}{\kappa}\alpha \dot{\gamma}(\phi-\phi_c).
    \label{eqn:Darcy-Reynolds}
\end{equation}

This equation represents a local constitutive law, which can then be extended to impact of a sphere into a saturated granular bed, with initial volume fraction $\phi_0$, using several more assumptions. During impact, there is a shearlike deformation that occurs beneath the intruder. The first important assumption is that this shear deformation occurs over a length scale $L$, which is proportional to the square root of the \textit{instantaneous} contact area between the intruder and the fluid-grain mixture. In this picture, $P_f$ then sets the pressure scale for a frictional intrusion-resisting force. This represents a qualitative difference between Darcy-Reynolds theory and recent theories that have been proposed to describe impact into dense suspensions, where propogating dynamically jammed fronts play a crucial role~\cite{waitukaitis2012impact,peters2014quasi,han2016high,brassard2021viscous}. Something like Darcy-Reynolds theory likely describes why these these dynamically jammed fronts remain solidified, due to very small (e.g., cornstach) particles with very low associated permeability. However, the theories describing each of these two systems cannot be directly applied to the other; we discuss further at the end of the Discussion section.

\subsection{\label{sec:L}The role of shear length scale $L$}
Assuming that the shear length scale $L$ is set by the square root of the contact area between the ball and the material, the shear rate is $\dot{\gamma}\sim v/L$, where $v$ is the speed of the intruder. The pore-pressure effects also act over length scale $L$, so $\nabla^2 P_f \sim P_f / L^2$. Thus, Eq.~\eqref{eqn:Darcy-Reynolds} becomes
\begin{equation}
    P_f \sim \frac{\eta_f}{d^2}\alpha L v (\phi-\phi_c),
    \label{eqn:Darcy-Reynolds-L}
\end{equation}
which sets the characteristic pressure on the sphere by the material. Assuming that Darcy-Reynolds pressure dominate all other forces, and that the intruder predominantly feels a frictional slowing down force, the equation of motion can be written as
\begin{equation}
    m \ddot{z} = - A \pi L^2 P_f,
    \label{eqn:eq-of-mot}
\end{equation}
where $m$ is the sphere mass, $z$ is the penetration depth ($v = \dot{z}$, $a = \ddot{z}$), and $A$ is an effective friction coefficient. The effective mass density $\rho_s$ can be defined as $\rho_s \equiv 6m / \pi D^3$, where $D$ is the sphere diameter. 

\subsection{Testable predictions from Darcy-Reynolds Theory}

If the Darcy-Reynolds pressure is very large, then the penetration depth of the sphere is small, i.e., $z\ll D$ (this is observed experimentally for small grains). In this case, $L^2 \approx Dz$ by a small-angle approximation. Combining, one obtains 
\begin{equation}
    \frac{\pi}{6}\rho_sD^3 \ddot{z} = -\frac{\pi A \alpha\eta_f D^{3/2}\Delta\phi}{d^2}z^{3/2}\dot{z},
    \label{eqn:eom-full}
\end{equation}
where $\Delta\phi = \phi-\phi_c$. After integrating in time, Eq.~\eqref{eqn:eom-full} yields a dimensionless equation of motion
\begin{equation}
    d\tilde{z}/d\tilde{t} = -(2/5)\tilde{z}^{5/2}+1,
    \label{eqn:dim-eom}
\end{equation}
where $\tilde{z} = z/Vt_m$, $\tilde{t}=t/t_m$, and 
\begin{equation}
    t_m = \frac{D}{V}\left(6A\alpha\frac{\eta_f D}{\rho_s d^2 V}\Delta\phi\right)^{-2/5},
    \label{eqn:t_m}
\end{equation}
The initial conditions are given by $z(0)=0$ and $\dot{z}(0) = V$, where $V$ is the initial velocity at impact. This means that the dimensionless velocity at impact is $\tilde{V} = 1$, since $d\tilde{z}/d\tilde{t} = \dot{z}/V$ and $\dot{z} = V$ at initial impact. 

Numerically solving Eq.~\eqref{eqn:eom-full} yields a deceleration-versus-time curve that rises, peaks at characteristic time set by $t_m$, and decreases. Such a curve can be seen in the Supplemental Material of Ref.~\cite{jerome2016} as well as in comparison to our experimental data in Fig.~\ref{fig:d-results-1}(b) (dashed line). The peak dimensionless acceleration $\tilde{a}_{\max} = a_{\rm max} t_m / V$ is therefore
\begin{equation}
    a_{\rm max} \propto \frac{V}{t_m} = \frac{V^2}{D}\left(6A\alpha\frac{\eta_f D}{\rho_s d^2 V}\Delta\phi\right)^{2/5}.
    \label{eqn:scaling-law}
\end{equation}
This equation predicts, for example, a peak force scaling via $a_{\rm max} \propto d^{-0.8} \eta_f^{0.4} V^{1.6}$, which can be explicitly tested. 

The derivation of Eq.~\eqref{eqn:scaling-law} contains several assumptions.  The breakdown of the validity of these assumptions leads to measurably different scaling behaviors. One assumption discussed by Jerome is that $\phi_c$ can be strain rate dependent, shifting the solidification response \emph{during} impact. Other deviations are possible; for example, in the case that the Darcy-Reynolds pressure is not sufficiently large to preserve $z\ll D$ throughout the impact, then Eq.~\eqref{eqn:eom-full} onward must be reevaluated. In the case where the Darcy-Reynolds pressure is dominant but still small enough that the penetration depth becomes similar to or larger than $D$, the contact area is simply proportional to the intruder cross sectional area, meaning that $L\approx D$, not $L\approx \sqrt{Dz}$. This means that Eq.~\eqref{eqn:eom-full} becomes
\begin{equation}
   \ddot{z} = -\frac{6  A \alpha\eta_f \Delta\phi}{\rho_s d^2}\dot{z}.
    \label{eqn:eom-full-mod}
\end{equation}
Thus, in this limit, the force on the impacting sphere is proportional to its speed, suggesting $a_{\rm max} \propto V$. Note that this linear dependence on velocity is reminiscent of Stokes drag, although the physical mechanism is different.

Equation~\eqref{eqn:eom-full-mod} predicts exponential decay in the velocity and acceleration, specifically
\begin{equation}
    \ddot{z} = -\frac{V}{\tau} e^{-t / \tau},
    \label{eqn:exp-decay}
\end{equation}
where $\tau = \frac{\rho_s d^2}{6  A \alpha\eta_f \Delta\phi}$. However, we note that during the initial stages of penetration, $L^2 \approx D z$ would still be valid, so we expect a buildup of the force before exponential decay takes over as the intruder passes through different scaling regimes. 

The equations in this section contain specific predictions about the dynamics, which are testable via experiments, as we show below. We note that there are other assumptions in these equations which may not always be valid, such as the assumption that $A$ is a constant that is independent of any system parameter. It is also reasonable to consider that the pore-pressure effects act over the particle scale $d$, giving $\nabla^2 P_f \sim P_f / d^2$ and hence additional scaling dynamics that we did not explore in detail. These and other assumptions may be responsible for deviations from these predictions, as we discuss below.


\section{\label{sec:methods}Experimental Methods}

To test the predictions from the previous section, we perform experiments of spherical intruders dropped into fluid-saturated granular beds. Since the dependence on $\Delta \phi$ was already confirmed in Ref.~\cite{jerome2016}, we vary other parameters, especially $V$, $d$, and $\eta_f$. We also vary $m$ and $D$, but over more modest range. We use five distinct sets of glass beads (Mo-Sci), with $d$ ranges of (1) 53-75, (2) 75-106, (3) 180-212, (4) 300-425, and (5) 600-850 in units of $\mu$m,. 
\subsection{\label{sec:twofluids}Details of the fluids}
We vary $\eta_f$ in two ways. First, we add glycerol to water in various concentrations, using data from~\cite{doi:10.1021/ie50501a040} to estimate the viscosity of the resulting mixture. By volume glycerol was diluted with water in 25\% increments and then converted to mPas for data analysis. Using these increments we achieve viscosity of 1~mPas (water), 2.4~mPas (25\% glycerol), 7.9~mPas (50\% glycerol), 24~mPas (75\% glycerol), and 1412~mPas (100\%). All viscosity experimental data was collected in an ambient temperature between 21.7-24.4~$^o$C, minimizing the impact of viscosity variance due to temperature variance.

\begin{figure}[!th]
    \raggedright (a) \hspace{52mm} (b) \hspace{52mm} (c) \\
    \centering
    \includegraphics[trim = 0mm 0mm 0mm 0mm, clip, width = 0.33\columnwidth]{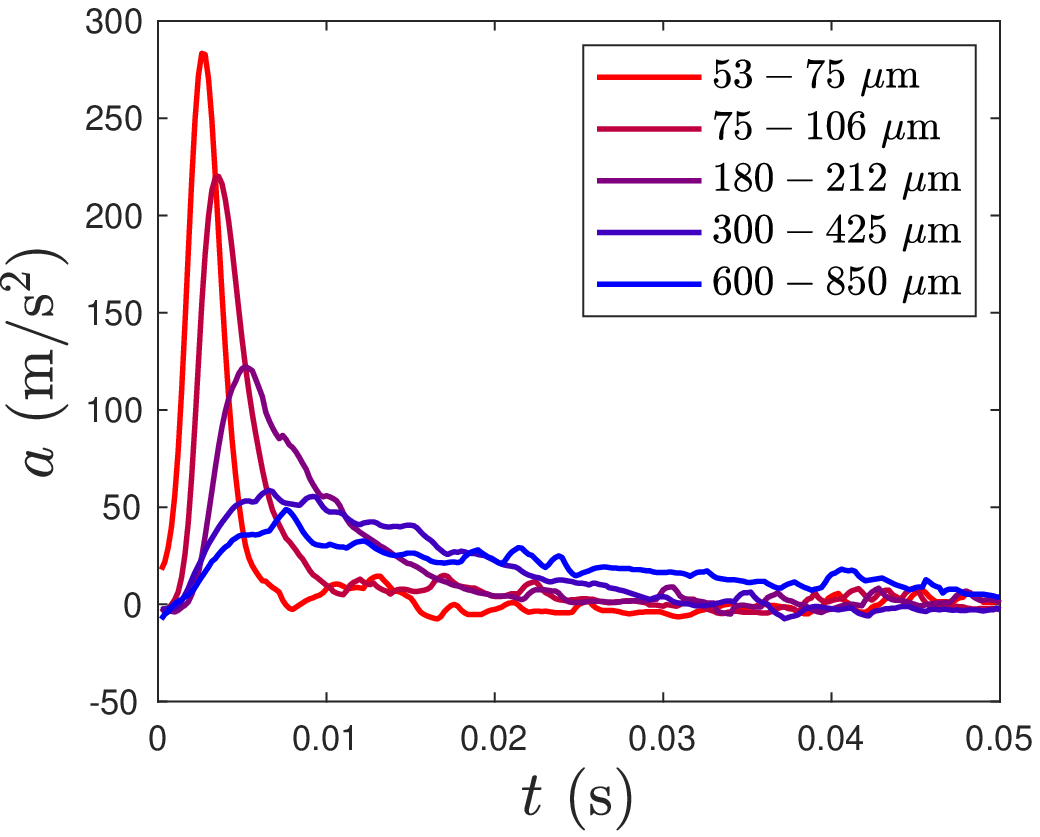}
    \includegraphics[trim = 0mm 0mm 0mm 0mm, clip, width=0.33\columnwidth]{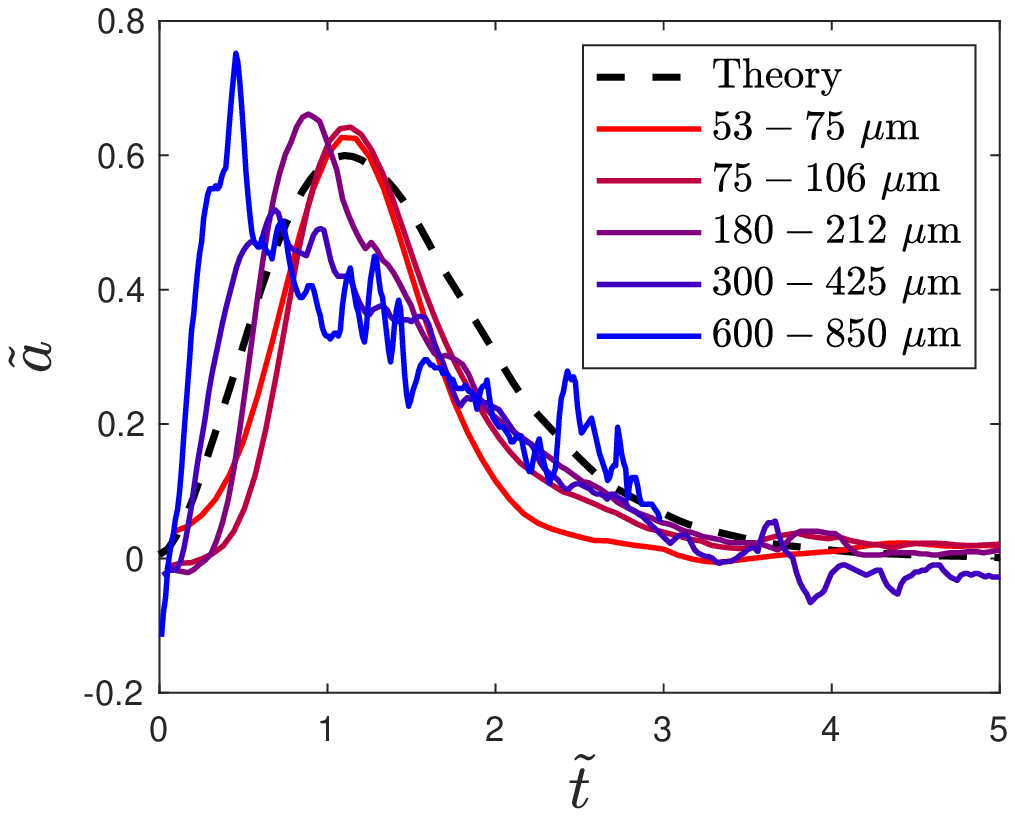}
    \includegraphics[trim = 0mm 0mm 0mm 0mm, clip, width=0.32\columnwidth]{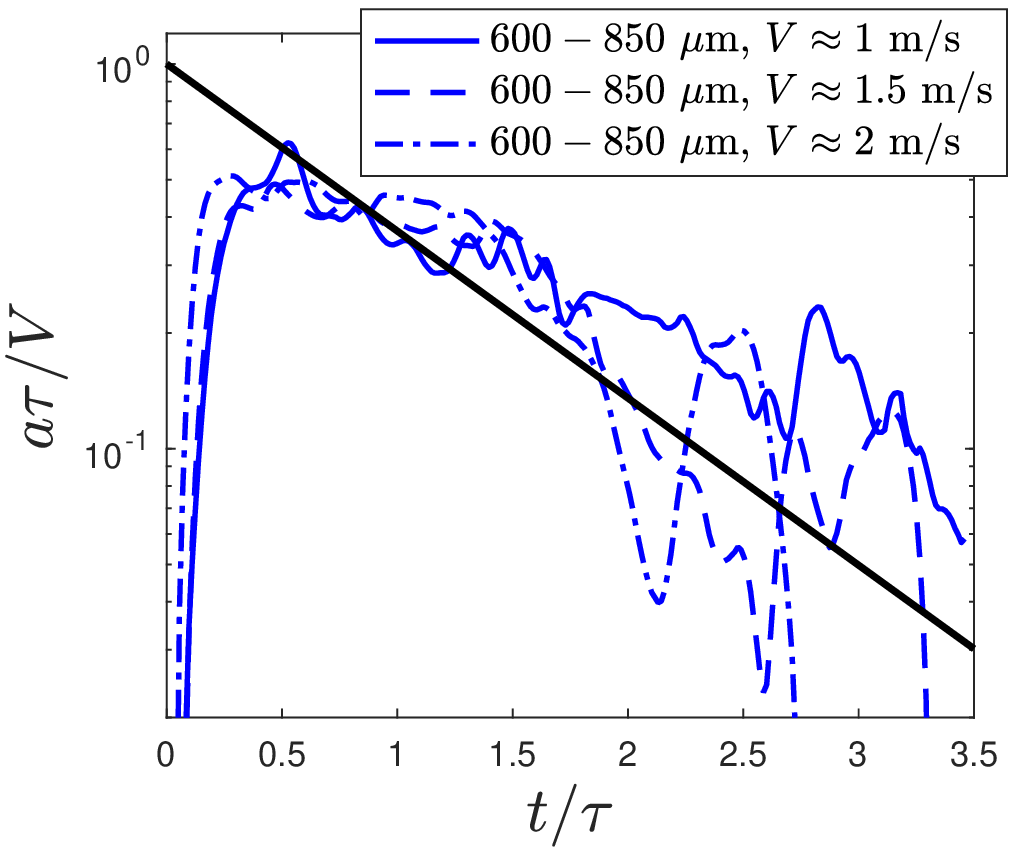}
    \caption{(a) Acceleration $a = -\ddot{z}$ as a function of time for impacts with similar impact speed, $V\approx 1$~m/s, for each of the five values of $d$. (b) Dimensionless acceleration $\tilde{a}$ as a function of dimensionless time $\tilde{t}$, as defined in Eq.~\eqref{eqn:dim-eom}. Smaller grains collapse well, where the assumptions leading up to this equation are valid. (c) Three representative curves of acceleration versus time for the largest beads, where the assumptions for Eq.~\eqref{eqn:dim-eom} are not valid. These larger grains are better captured by Eq.~\eqref{eqn:exp-decay}; see text for discussion.}
    \label{fig:d-results-1}
\end{figure}

Second, we use Ferrotec EFH1 Ferrofluid to conduct similar test to the glycerol, where we adjust the viscosity with an external magnetic field. The ferrofluid has a base viscosity of 6~mPas ~\cite{ferrofluid_2018}, and we increase the viscosity by adding an external magnetic field using arrays of permanent magnets as well as a large solenoid for smaller adjustments. We measure the magnetic field using a Hall probe, and we verify that it is fairly constant over the volume where the impact will take place (less than 50\% variation). Magnetic readings were taken from edge to edge in the container to ensure symmetric magnetic forces on each side of the impact zone.  We characterize the profile by a single number, $B_0$, corresponding to the magnetic field at the point where the impact will take place. We take data for $B_0=0$ mT (no field), 3~mT, 9~mT, 14~mT, and 31~mT. Based on ~\cite{patel2003viscosity}, we estimate that viscosity will increase by roughly two orders of magnitude over this range of $B_0$. We return to this point below in Sec.~\ref{sec:results}, when we describe results from impacts into ferrofluid-saturated beds. We used SEM to identify the interaction between the beads and ferrofluid, as shown in Fig.~\ref{fig:sketch}(d). After drying the silica beads and ferrofluid mixture SEM images show that the silica beads are covered in ferrofluid particles. Importantly, there are characteristic fluid contact marks left where the ferrofluid collected at the contact points while drying. The ferrofluid appears to have maintained fluid like properties depositing the nanometer sized iron oxide particles as it dried which suggests that the ferrofluid retains its fluid-like properties also on a scale much smaller than $d$. There are also no large cohesive-looking ferrofluid particle agglomerations present that would interfere with direct solid-solid silica bead contacts. 

\subsection{\label{sec:sampleprep}Sample Preparation}
Samples are prepared by filling a 8x8~cm acrylic container with an open top with the relevant fluid to a level of roughly 10 cm. We then slowly add particles of given size until the particle packing level is at approximately the same height as the fluid. By weighing both the fluid and the grains as they are poured, we can estimate the volume ratio and thus packing fraction of the suspension. We observe $\phi \approx 0.59$ for all samples with the exception the largest beads ($d = 600$ to $850$~$\mu$m), where $\phi \approx 0.60$ is observed. We assume that $\phi>\phi_c$, which is corroborated by the fact that we observe a strong solidification during impact, consistent with the results for $\phi>\phi_c$ in Ref.~\cite{jerome2016}.

Our sample preparation protocol depends on the fluid used. For water-based suspensions, we stir the suspension after impact and let it settle. For the more viscous samples stirring is challenging to do without introducing bubbles. After each impact the material furthest from the impact is moved to fill the impact crater and we wait for it to settle. Because the more viscous samples did not settle when completely stirred, settling experiments were done on to ensure that $\phi$ does not change significantly from run to run via the application of the protocol. Visual inspection shows that $\phi$ did not change in any significant manner until after many hours to days of settling. Because all the viscosity impact data was collected in less than four hours we minimize the risk of $\phi$ changing due to settling. We further verify that $\phi$ is not playing a role in our measurements by the fact that there is no significant change in the small fluid layer at the top of the sample. We also perform multiple sets of experiments with the same sample, and we find that our measurements are repeatable.

\subsection{\label{sec:impacts}Generating impacts}
Steel spheres are fixed to the end of threaded rods and dropped from various heights $H$ by releasing them using an electromagnet. Impact velocities are inferred by $V = \sqrt{2gH}$, where $g$ is the gravitational acceleration; we also confirm these velocities using high-speed video for selected cases. We measure the acceleration using an onboard accelerometer (Sparkfun with ADXL377). We vary $m$ and $D$ by adding mass or using different diameter steel spheres. A sketch of the setup is shown in Fig.~\ref{fig:sketch}(a).


\section{\label{sec:results}Experimental Results}


\subsection{Phenomenology}
Figure~\ref{fig:d-results-1}(a) shows $a(t)$ for five impacts, one for each range of $d$, all with impact velocity $V\approx 1.1\pm 0.1$~m/s. These curves clearly demonstrate that, consistent with the Darcy-Reynolds picture, smaller $d$ leads to more sharply peaked deceleration profiles with larger values of $a_{\rm max}$. Figure~\ref{fig:d-results-1}(b) shows $-\tilde{a}$, where $\tilde{a} \equiv d^2 \tilde{z} / d\tilde{t}^2$, plotted as a function of $\tilde{t}$ for each of the five curves shown in Fig.~\ref{fig:d-results-1}(a). For the three smallest particle size ranges ($d = 53$ -- 75 $\mu$m, $d = 75$ -- 106 $\mu$m, and $d = 180$ -- 212 $\mu$m), the Darcy-Reynolds pressure is sufficiently large that $z\ll D$ is satisfied during the bulk of the trajectory, meaning that Eq.~\eqref{eqn:scaling-law} is valid. The rescaled experimental results for the samller agree well with a numerical solution of Eq.~\eqref{eqn:dim-eom}, which is shown as a black dashed line. 


To rescale the experimental data in Fig.~~\ref{fig:d-results-1}(b), we need to set $t_m$ for each curve. Several parameters in $t_m$ are not directly measurable or known \textit{a priori}, but a suitable collapse (at least for small $d$) is found by setting $6 A \alpha \Delta\phi = 80$. The packing fraction differential $\Delta \phi$ is of order 10$^{-2}$, meaning that the product $A\alpha$ is of order $10^3$. Above, we estimated $\alpha \approx 20$, meaning that the effective friction coefficient $A \approx 50$. While this value appears very high for a friction coefficient, we note that it is not a simple friction coefficient and that a previous study on granular intrusion found a value of approximately 35 for a similar parameter~\cite{Brzinski2013}. Additionally, we note that the assumptions leading up to Eq.~\eqref{eqn:Darcy-Reynolds-L} may affect the value of $\alpha$. In any case, the quantities are within reasonable physical bounds, and assuming $6 A \alpha \Delta\phi = 80$ yields good agreement between the experimental trajectories and the theoretical prediction of Eq.~\eqref{eqn:scaling-law} for small $d$.

In contrast, for the two largest particle sizes, the Darcy-Reynolds pressure is much smaller, meaning that the sphere is able to penetrate more deeply into the material. In this case, $z \ll D$ is not satisfied, so it is not surprising from the theoretical prediction that arises from solving Eq.~\eqref{eqn:dim-eom} numerically that the experimental results do not match the Jerome predictions. This argument can be made quantitative by estimating the penetration depth $z^*$ at peak deceleration in the following way. By assuming that the average deceleration between $t=0$ and the time $t_{\rm max}$ corresponding to the peak deceleration is equal to half of the peak value (corresponding to approximating the rise in deceleration as linear), then $z^* \approx Vt_{\rm max} - a_{\rm max}t_{\rm max}^2/4$. We can then use $z^*/D$ as a dimensionless measure of how far the sphere has penetrated relative to its own diameter at $t_{\rm max}$. For the smallest particles, with $d = 53$ -- 75 $\mu$m, we find typical values of $z^*/D \sim 0.1$, meaning that $z\ll D$ is a reasonable approximation. For the largest particles, with $d = 600$ -- 850 $\mu$m, we find typical values of $z^*/D \sim 1$ or larger.

The data for $d = 600$ -- 850 $\mu$m in particular appears to be better described by a sharp rise followed by quasi-exponential decay, consistent with Eq.~\eqref{eqn:exp-decay}. Figure~\ref{fig:d-results-1}(c) shows acceleration versus time for three impacts into beds with $d = 600$ -- 850 $\mu$m and with initial velocities $V \approx 1$, 1.5, and 2~m/s; the data for $V \approx 1$ is the same curve shown in panel (a). Note that the axes are rescaled using $\tau = \frac{\rho_s d^2}{6  A \alpha\eta_f \Delta\phi}$, as defined in conjunction with Eq.~\eqref{eqn:exp-decay}, as well as the impact velocity $V$ on the horizontal axis. The solid black line shows the theoretical prediction from Eq.~\eqref{eqn:exp-decay}. Note that the values used in the rescaling here are fully determined by the collapse in panel (b), using $6 A \alpha \Delta\phi = 80$, leaving no free parameters.

\begin{figure}
    \raggedright \hspace{2mm} (a) \hspace{52mm} (b) \hspace{52mm} (c) \\
    \centering

    \includegraphics[trim = 0mm 0mm 0mm 0mm, clip, width = 0.32\columnwidth]{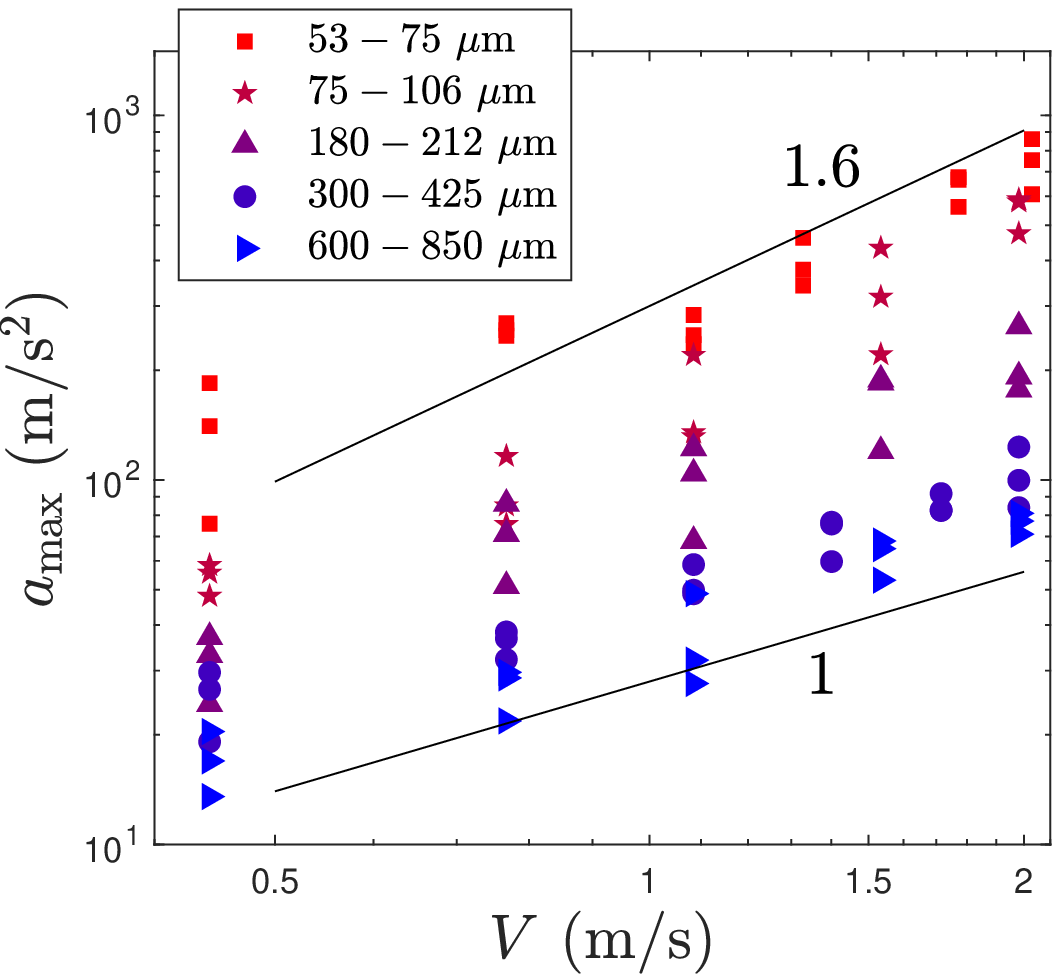}
    \includegraphics[trim = 0mm 0mm 0mm 0mm, clip, width = 0.32\columnwidth]{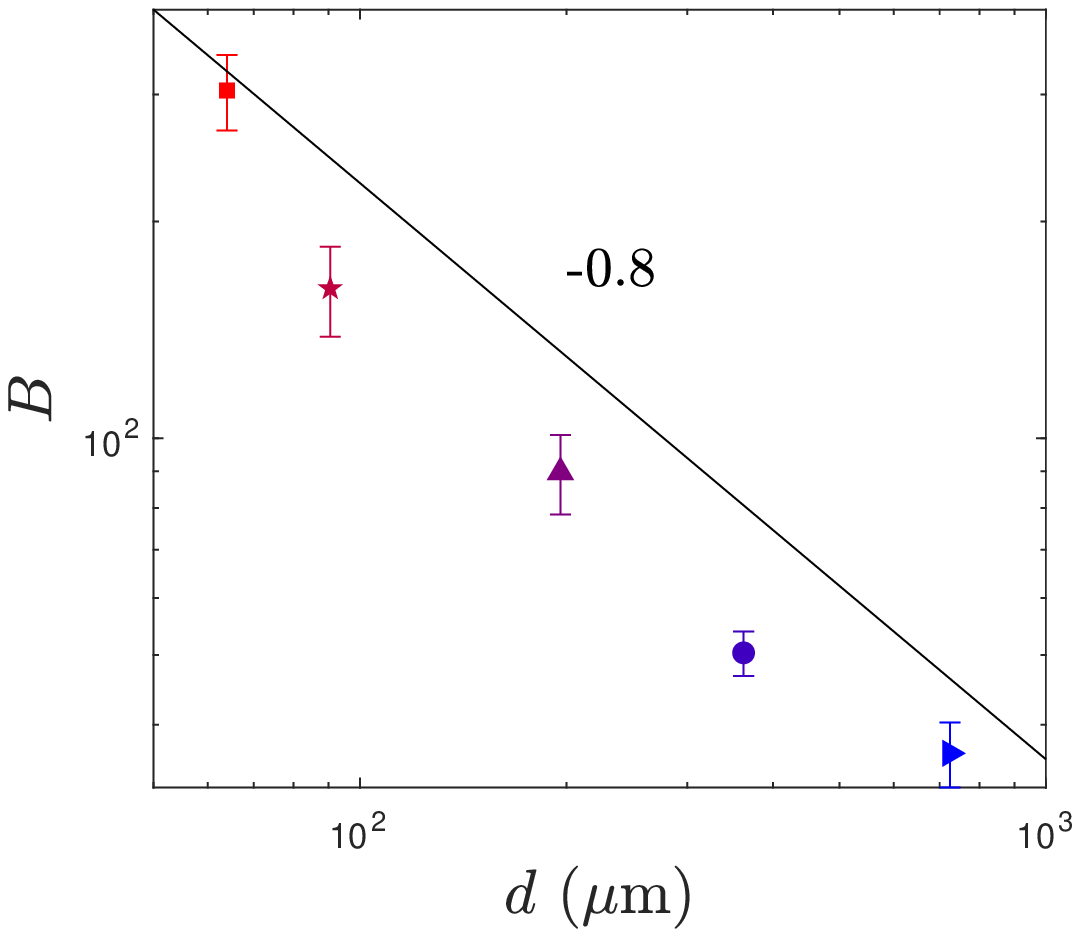}
    \includegraphics[trim = 0mm 0mm 0mm 0mm, clip, width = 0.32\columnwidth]{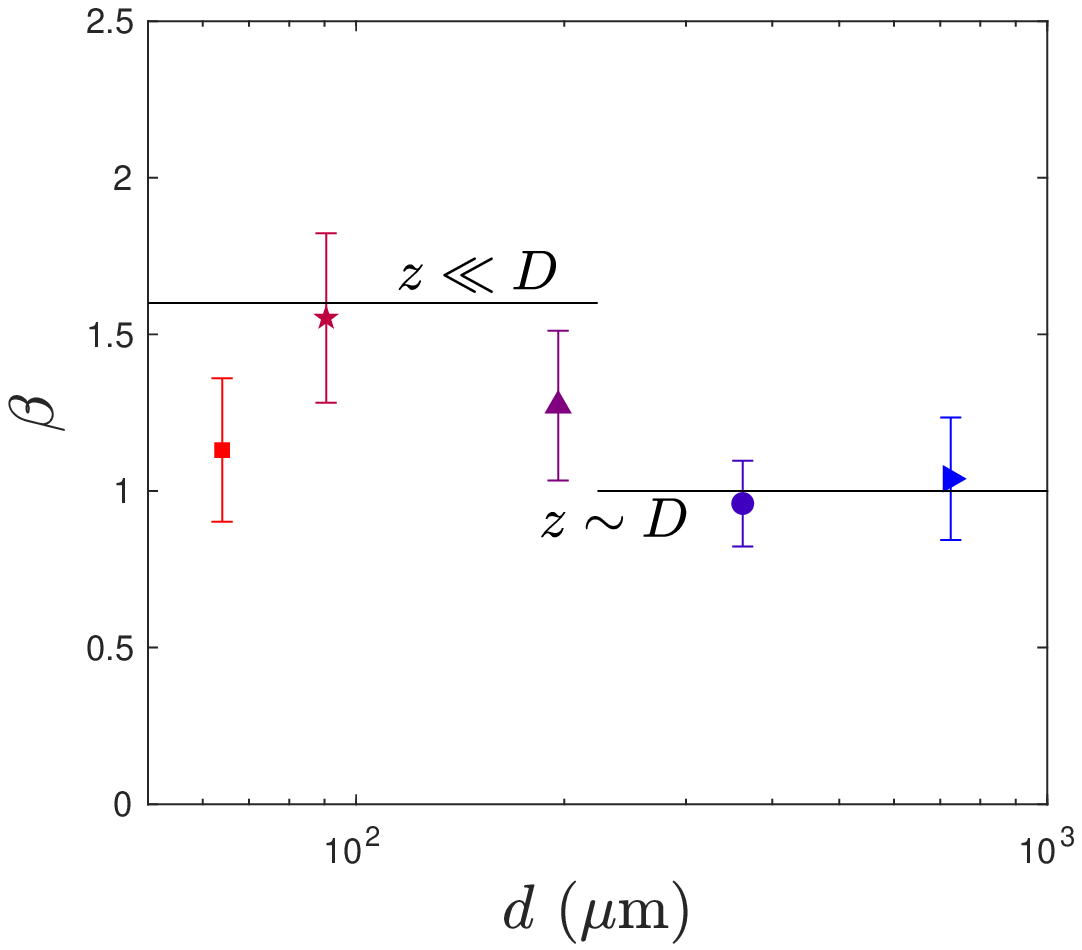}     
    \caption{(a) $a_{\rm max}$ versus $V$ for all five values of $d$. Smaller beads obey the scaling law in Eq.~\eqref{eqn:scaling-law}, as expected, since $z\ll D$ for these impacts. Larger beads obey $a_{\rm max} \propto V$ from Eq.~\eqref{eqn:exp-decay}, as expected, since $z\sim D$ for these impacts; see text for discussion. (b,c) We perform linear fits to the logarithmic data in panel (a) to obtain the best fit for the function $a_{\rm max} = BV^\beta$. DRT predictions are shown: $B \propto d^{-0.8}$ for panel (b); $\beta = 1.6$ for small beads and $\beta = 1$ for large beads in panel (c).}
    \label{fig:d-results-2}
\end{figure}

\subsection{Scaling results for $V$ and $d$}

Figure~\ref{fig:d-results-2}(a) shows $a_{\rm max}$ versus $V$ for all five particle sizes. Equation~\eqref{eqn:scaling-law} shows that DRT predicts $a_{\rm max} \propto V^{1.6}$, under several assumptions, including $z\ll D$. Experimental results are more consistent with this prediction for smaller $d$ (red squares, green stars, and black triangles) and larger $V$ where these assumptions are valid. For large $d$ (e.g., blue circles and magenta triangles), we observe $a_{\rm max}\propto V$, as predicted by Eq.~\eqref{eqn:exp-decay} and consistent with the collapse in Fig.~\ref{fig:d-results-1}(c). Figure~\ref{fig:d-results-2}(b) and (c) show the result of fits to the $a_{\rm max}$-versus-$V$ data in Fig.~\ref{fig:d-results-2}(a) of the form $a_{\rm max} = B V^\beta$ (error bars represent 95\% confidence intervals). We find good agreement with the prediction from Eq.~\ref{eqn:scaling-law} that $B \propto d^{-0.8}$. The best fit for the exponent $\beta$ is consistently smaller than the prediction of $\beta\approx 1.6$ from Eq.~\eqref{eqn:scaling-law}, even for small $d$ where we would expect it to be valid. However, we note that the low value of $\beta$ for the smallest particles seems to be caused by outliers at low $V$; the large-$V$ data appears very consistent with $\beta \approx 1.6$. Overall, $a_{\rm max} \propto V^{1.6}$ appears to be highly consistent with the data at small $d$. For larger $d$, where the Darcy-Reynolds pressure is smaller and we expect Eq.~\eqref{eqn:exp-decay} to be applicable, we find $a_{\rm max} \propto V$ as expected. Overall, our data for varied $d$ agree well with the predictions of Darcy-Reynolds theory.

\begin{figure}[!b]
    \raggedright \hspace{2mm} (a) \hspace{82mm} (b) \\
    \centering
    \includegraphics[trim = 0mm 0mm 0mm 5mm, clip, width = 0.49\columnwidth]{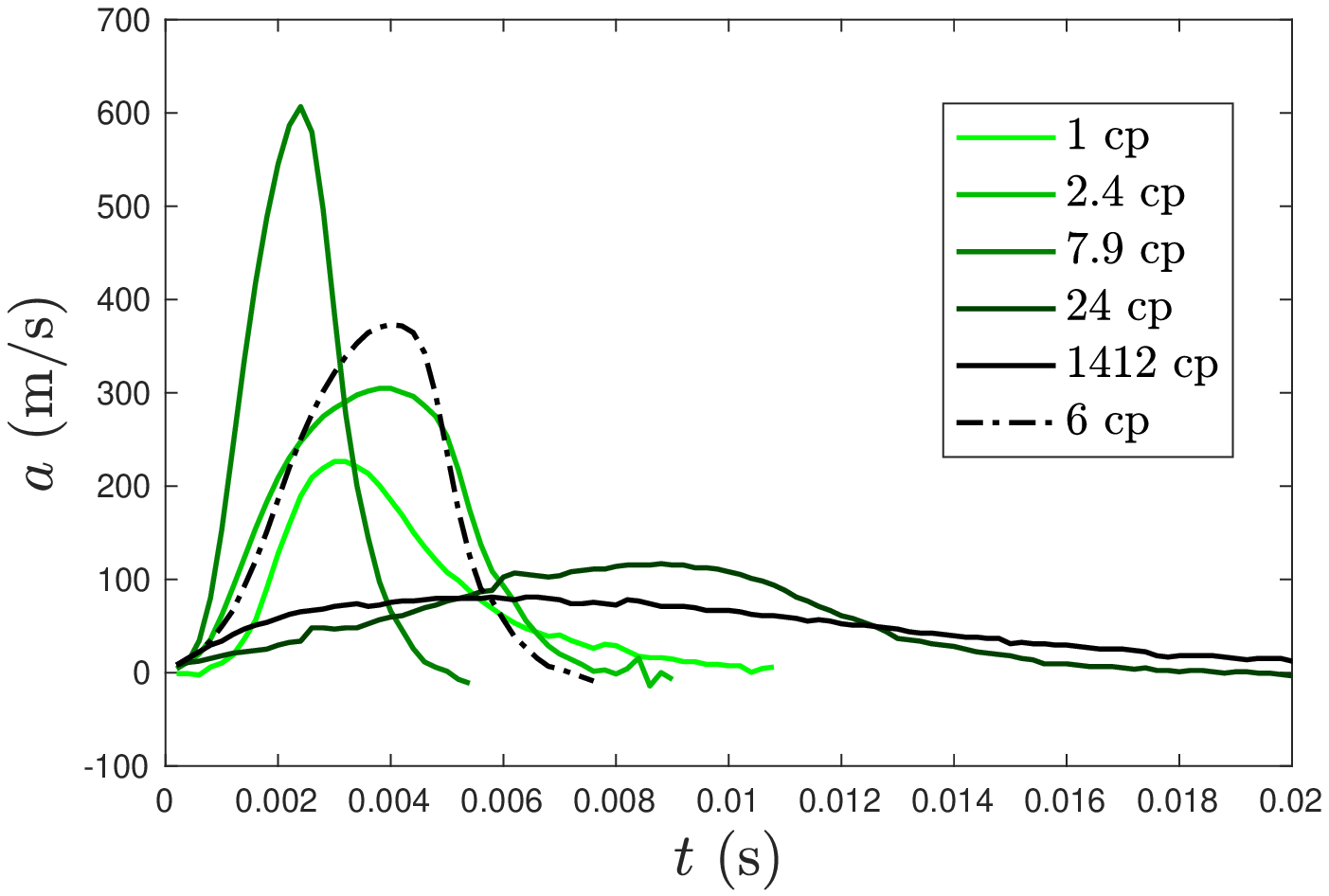}
    \includegraphics[trim = 0mm 0mm 0mm 5mm, clip, width = 0.49\columnwidth]{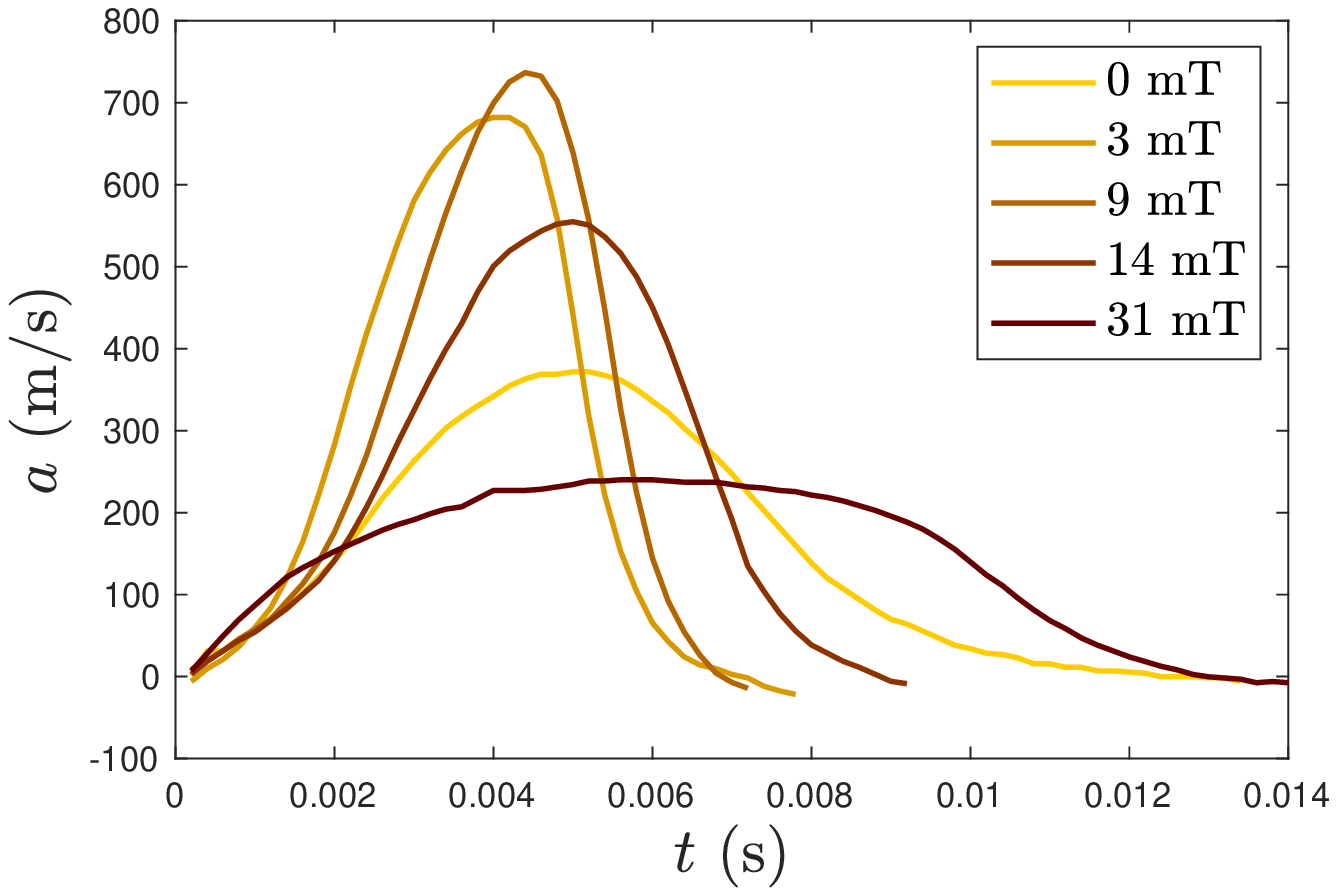}
    \caption{Acceleration vs time for different impacts into beds with $d = 75$ -- 106 $\mu$m and varying fluid viscosity. Panel (a) shows impacts with $V\approx 1$ where $\eta_f$ is varied by adding glycerol to water. Panel (b) shows impacts with $V \approx 2.5$ where $\eta_f$ is varied by using a ferrofluid and changing the external magnetic field. A zero-field ferrofluid impact result is also shown in panel (a) for the ferrofluid (dash-dotted line).}
    \label{fig:visc-results-1}
\end{figure}

\subsection{Viscosity dependence}
Turning now to the impact behavior for varied $\eta_f$, Fig.~\ref{fig:visc-results-1} shows results for impacts at varying $V$ and $\eta_f$ with constant particle size $d = 75$-106~$\mu$m. Fig.~\ref{fig:visc-results-1}(a) shows typical acceleration curves with similar $V\approx 1.2$ but with varied $\eta_f$. The solid curves represent water-glycerol mixtures, and the dash-dotted curve (6 cp) represents the ferrofluid with no applied magnetic field. These curves appear qualitatively similar to those in Fig.~\ref{fig:d-results-1}. However, in contrast with the predictions of DRT, the peak value $a_{\rm max}$ shows non-monotonic behavior as $\eta_f$ is increased: $a_{\rm max}$ increases with $\eta_f$ up to $\eta_f = 7.9$ cp but then decreases dramatically as $\eta_f$ is further increased. This decrease in $a_{\rm max}$ is qualitatively inconsistent with DRT.

\begin{figure}[!t]
   \raggedright \hspace{2mm} (a) \hspace{52mm} (b) \hspace{52mm} (c) \\
    \centering
    \includegraphics[trim = 0mm 0mm 0mm 0mm, clip, width = 0.32\columnwidth]{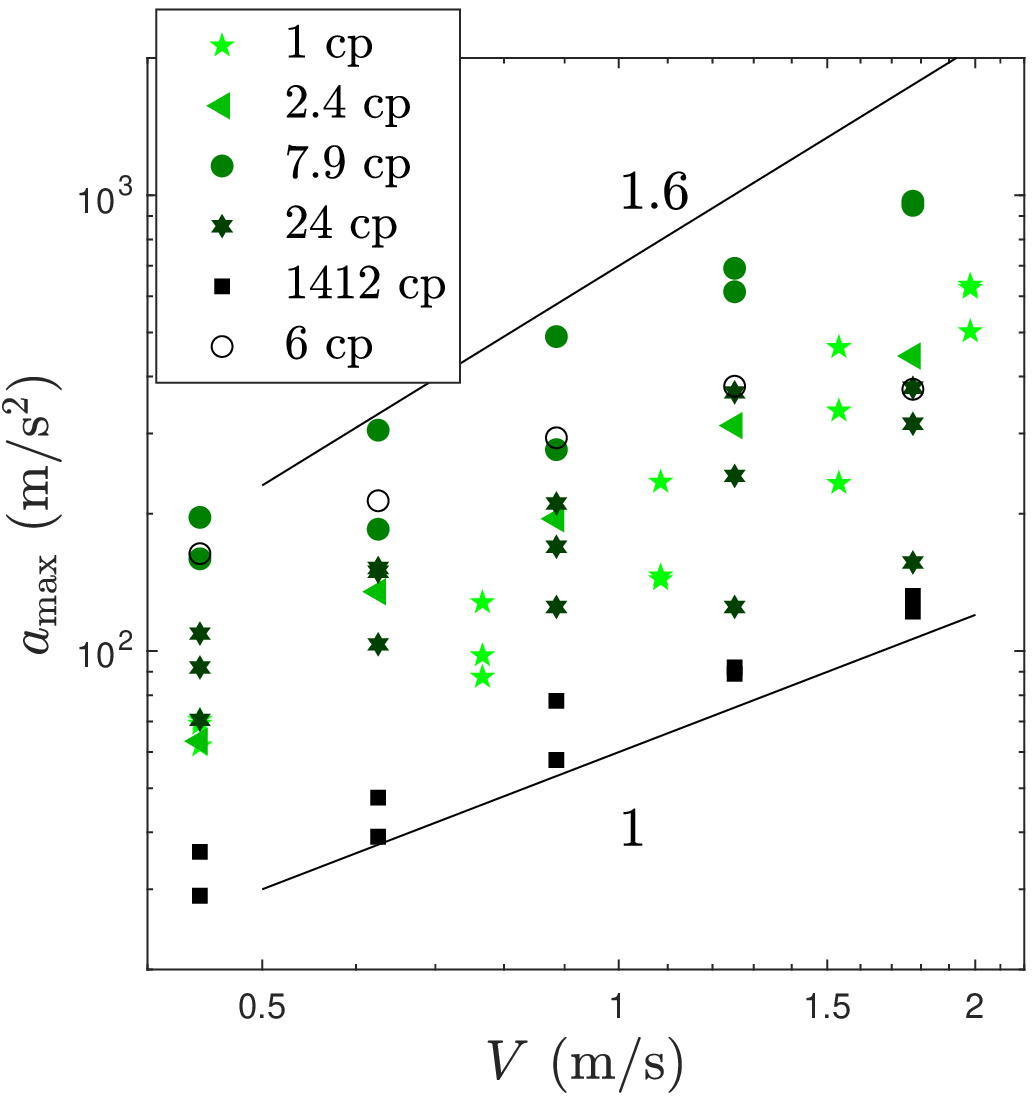}
    \includegraphics[trim = 0mm 0mm 0mm 0mm, clip, width = 0.32\columnwidth]{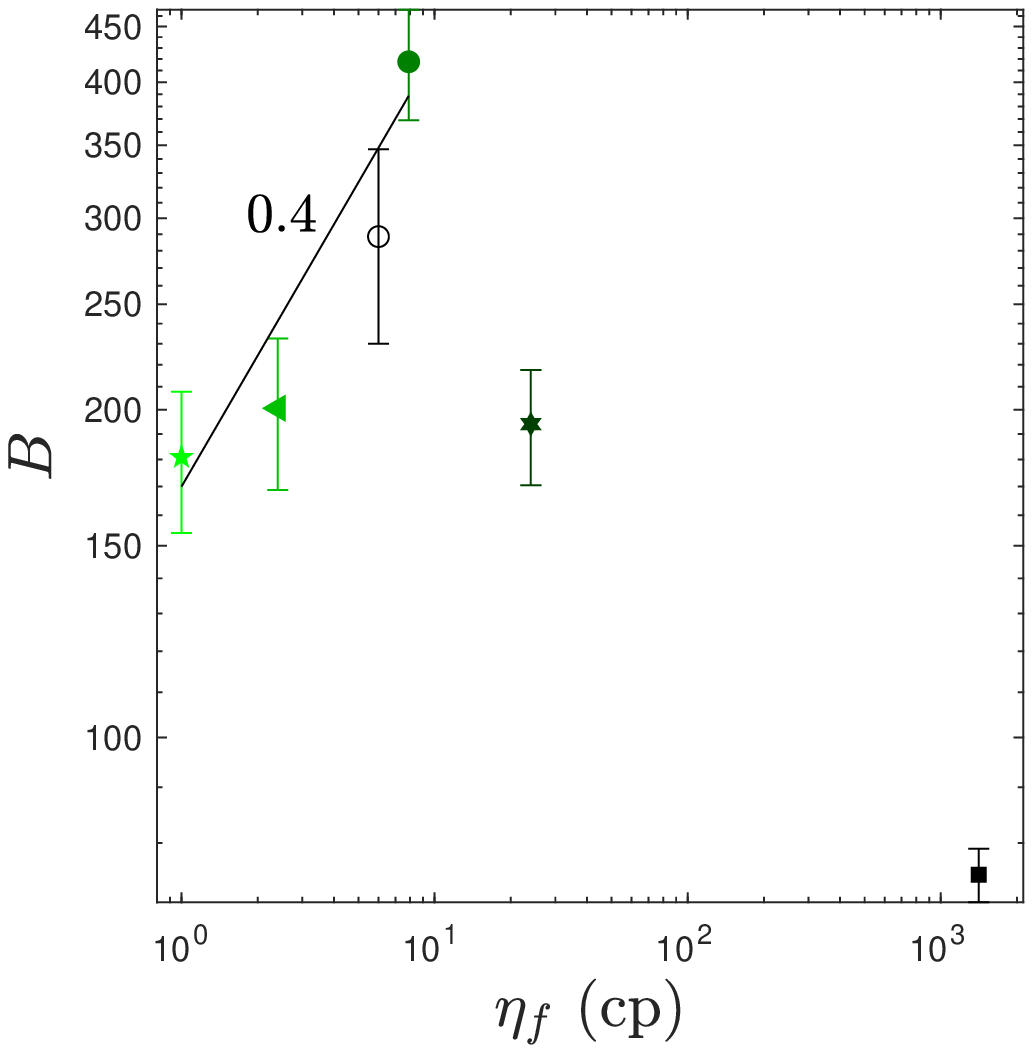}
    \includegraphics[trim = 0mm 0mm 0mm 0mm, clip, width = 0.32\columnwidth]{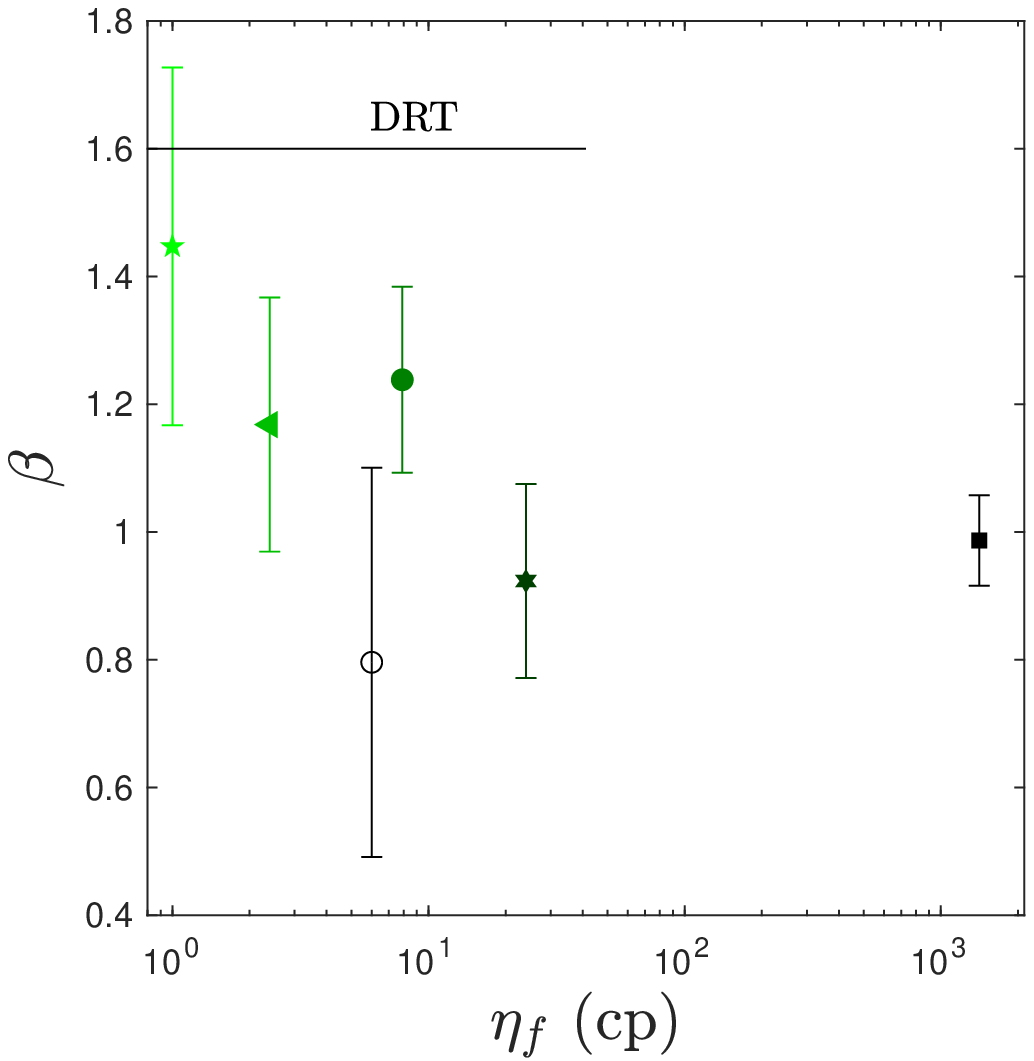}     
    \caption{(a) $a_{\rm max}$ versus $V$ for all values of $\eta_f$. (b,c) We perform linear fits to the logarithmic data in panel (a) to obtain the best fit for the function $a_{\rm max} = BV^\beta$. DRT predictions are shown: $B \propto \eta_f^{0.4}$ for panel (b); $\beta = 1.6$ in panel (c). Similar to the data from Fig.~\ref{fig:visc-results-1}, we find that DRT breaks down for $\eta_f>10$~cp, and forces begin to decrease with increasing viscosity.}
    \label{fig:visc-results-2}
\end{figure}

As stated in Sec.~\ref{sec:intro}, one motivation of these experiments was to move toward tunable particle-fluid mixtures. In particular, if the viscous fluid were replaced with a ferrofluid, then the particle-fluid mixture could be strengthened \textit{in situ} by applying an external magnetic field. Our results for glycerol-water mixtures, shown in Fig.~\ref{fig:visc-results-1}(a), suggest that this may not be the case. Figure~\ref{fig:visc-results-1}(b) shows that the breakdown of thickening at higher viscosities reproduces when the viscous fluid is no longer glycerol-water mixtures but is instead replaced by a ferrofluid. The initial viscosity of the ferrofluid is $\eta_f=6$~cp, which is just below the value of $\eta_f\approx 10$~cp where we expect to see the largest values of $a_{\rm max}$ for the range of $V$ we study here. Therefore, we expect that as the applied magnetic field is increased, we should see a slight increase in $a_{\rm max}$ followed by a decline. Figure~\ref{fig:visc-results-1}(b) shows five curves of $a(t)$ with similar $V\approx 2.5$~m/s with five different magnetic field strengths, $B_0 = 0$, 3, 9, 14, and 31~mT. We again find that $a_{\rm max}$ increases with $\eta_f$ (which we vary indirectly through $B_0$) and then begins to decrease. Importantly, we find this result consistently for all values of $V$. Figure~\ref{fig:visc-results-2}(a) shows data for $a_{\rm max}$ versus $V$. The data for $\eta_f = 1$~cp (water), marked with green stars, are the same data shown with deep red stars in Fig.~\ref{fig:d-results-2}. Figure~\ref{fig:visc-results-2}(b) and (c) show results of best fits to these data in the form $a_{\rm max} = B V^\beta$. We again find that best fits for $\beta$ tend to be slightly smaller than the prediction of $\beta = 1.6$ from Eq.~\eqref{eqn:scaling-law}. The best fits for $B$ increase in a way that is consistent with the prediction of $B \propto \eta_f^{0.4}$ from Eq.~\eqref{eqn:scaling-law} for 1 cp (water), 2.4 cp (25\% glycerol), 6 cp (zero-field ferrofluid) and 7.9 cp (50\% glycerol). However, for larger values of $\eta_f$, we observe that $B$ begins to decrease, consistent with Fig.~\ref{fig:visc-results-1}.

\section{\label{sec:discussion}Discussion}

Here we have followed~\citet{jerome2016} to derive equations describing the dynamics of a sphere impacting a fluid-saturated granular bed, where the granular phase is compacted above the critical volume fraction such that it dilates under shear. The dilation (Reynolds) caused by the impact forces fluid to flow into the expanding pore structure, and the resulting Darcy pressure dominates the force on the intruder. We have expanded on the derivation from Ref.~\cite{jerome2016} to include the case where the sphere penetration depth becomes similar to or larger than its diameter.

The predictions from this theory were experimentally confirmed by~\citet{jerome2016} with regard to the dependence on $\Delta \phi$. Here, we performed additional experiments to test the theory's predictions on other parameters, specifically impact velocity $V$, particle size $d$, and fluid viscosity $\eta_f$. Our experimental results confirm the predictions of DRT for variation in $d$ over more than an order of magnitude. For small $d$, DRT as formulated in Ref.~\cite{jerome2016} works well. For larger $d$, a key approximation ($z\ll D$) for the context of impact dynamics is no longer applicable, so the equation of motion describing the impacting sphere's dynamics must be modified. However, with this modification, DRT still captures the observed behavior over a large range of $d$. When we vary $\eta_f$, we observe good agreement with DRT for $\eta_f$ between 1 and 10~cp, specifically that $a_{\rm max}\propto \eta_f^{0.4}$. However, for $\eta_f>10$~cp, we observe that forces generated during impact begin to decrease as $\eta_f$ is further increased. We validate this result with two methods: varying $\eta$ by adding glycerol to water as well as by using a ferrofluid with an externally applied magnetic field. Both approaches show that the forces generated during impact begin to decrease with increasing $\eta_f$ at $\eta_f\approx 10$~cp. This is qualitatively inconsistent with DRT, suggesting that a new theory is required. One possibility is that the frictional behavior of the grain-grain contacts changes drastically with a very viscous fluid (recall, a constant effective friction coefficient $A$ was assumed for all $d$ and $\eta_f$). For large $\eta_f$, grains may not be able to squeeze out the fluid to make frictional, solid-solid contact. A similar mechanism is often invoked to explain shear thickening behavior in dense suspensions~\cite{cates}.

Finally, as pointed out in~\cite{jerome2016}, the pore-pressure effects from DRT could help explain the dramatic response of impact into shear-thickening suspensions. Shear-thickening suspensions differ from saturated granular beds, in that the particles are not making solid-solid contact in the absence of driving. However, in these systems, $\phi_c$ decreases as stress is applied, causing dilation and thereby inducing a large Darcy pressure that solidifies the material. We emphasize that there are some key differences between the system discussed here and impact into dense suspensions, such as the fact that propagation phenomena play a key role in the impact response of shear-thickening suspensions but are not considered in DRT. However, DRT likely plays an important role in impact behavior into shear-thickening suspensions, primarily by holding together the dynamic solidlike region often observed in shear-thickening impact experiments.



\section{Acknowledgements}
We acknowledge funding from the Office of Naval Research under Grant No. N0001419WX01519 and by the Office of Naval Research Global Visiting Scientist Program VSP 19-7-001. We thank Drago Grbovic for taking SEM images.

\bibliography{references}
\end{document}